\documentclass[twocolumn, showpacs]{revtex4}
\usepackage{graphicx}
\usepackage{amsmath}
\usepackage{mathrsfs}
\usepackage{amsfonts}
\usepackage{overpic}
\usepackage{hyperref}
\usepackage{rotating}
\newcommand{\ket}[1]{|#1\rangle}
\newcommand{\bra}[1]{\langle #1|}

\begin{document}
\title{A Study on the Sudden Death of Entanglement}
\author{H. T. Cui\footnote{E-mail: cuiht@student.dlut.edu.cn}, K. Li, X. X. Yi}
\affiliation{Department of Physics, Dalian University of Technology,
Dalian 116024, China}
\date{\today}
\begin{abstract}
The dynamics of entanglement and the phenomenon of entanglement
sudden death (ESD) \cite{yu}  are discussed in bipartite systems,
measured by Wootters Concurrence. Our calculation shows that ESD
appears whenever the system is open or closed and is dependent on
the initial condition. The relation of the evolution of entanglement
and energy transfer between the system and its surroundings is also
studied.
\end{abstract}
 \maketitle

\section{introduction}
The dynamics of entanglement in bipartite systems has received great
attention since the work of Yu and Eberly \cite{yu1}, in which the
entanglement between the two particles coupled with two independent
environments became completely vanishing in a finite time. This
surprising phenomenon, contrary to intuition based on experience
about qubit decoherence,  intrigues great interests \cite{roszak,
ficek, liu, my, yu2, yu3}. Different from the original work of Yu
and Eberly \cite{yu1}, in which two independent particles embedded
into its own dissipative environment and there is no any direct or
indirect interaction, the effects of interaction between the
particles and the couplings to the same environment have been
discussed extensively in Ref. \cite{ficek} and \cite{liu}. In
\cite{ficek}, the authors show that for a special initial state the
entanglement disappeared in a finite time and then revived after a
dark period because of the interaction between the particles.
Furthermore the authors show that the entanglement sudden death
(ESD) is sensitive to the initial condition, as proved in
\cite{liu}. An important character in \cite{ficek, liu} is that the
environment is dissipative and the transfer of energy between the
system and environment is inevitable. Another important situation is
the dephasing environment, in which energy transfer from the system
to the environment does not occur. Some works have been devoted to
this issue \cite{roszak, yu2}. In \cite{roszak}, the authors show
that disentanglement is dependent on the initial condition and
temperature of the enviroment.

Although the extensive progress in understanding the
disentanglement, it is still unclear what reason causes this
phenomenon and what is the physics behind. In this paper we try to
give an enlightening discussion by examining some examples. In the
previous studies, the couplings with the environment has been
considered as the critical point of  disentanglement and ESD.
However, as will find in this paper, ESD can also happen in closed
systems. In our own point, the phenomena of disentanglement and ESD
stem from the dissipative terms in the Hamiltonian, independent of
the coupling with environment, and are sensitive to the initial
conditions. Moreover the revival of entanglement can also be
explained as the effect of the backaction terms in the Hamiltonian.

The paper is organized as follows. In Sec. II, we first review the
dynamics of entanglement  in (a) Tavis-Cumming model \cite{tc} and
(b) dephasing system respectively. In Sec. III, the dynamics of
entanglement in closed bipartite system has been discussed, and ESD
and revivals of entanglement have also been observed in this system.
The further discussions and conclusions is presented in the final
section.

\section{disentanglement in open systems}
The dynamics of entanglement in open systems has been discussed
extensively \cite{yu, yu1, ficek, liu, my, yu2}. However the physics
behind the evolution of entanglement is rarely touched. In this
section we try to present an physical interpretation for concurrence
by examining two examples. The first is the Tavis-Cumming model
\cite{tc}, in which two non-interaction qubits couple with the same
quantized field under the rotating-wave approximation. Another is
the dephasing model, in which two qubits embed into a multimode
quantized field and the interaction between the two qubits is also
considered.

\subsection{Tavis-Cumming Model}
For two spin-1/2 particles coupled with a single-mode cavity field,
the Hamiltonian is written as \cite{tc},
\begin{equation}\label{hjc}
H=\frac{\omega_0}{2}(\sigma_A^z + \sigma_B^z)+ \omega a^{\dagger}a +
g\sum_{i=A, B}(a\sigma_i^+ + a^{\dagger}\sigma_i^-),
\end{equation}
where $a^{(\dagger)}$ is bosonic operator and $\sigma_i^{\pm}$ are
the rising and lowing operator for spin-1/2. We focus our discussion
on the resonating case ($\omega_0=\omega$), in that the spectrum of
Hamiltonian can be obtained exactly.  For the whole system (system +
field), the evolution of the whole system is characterized by the
interacion between the system and field. However from the point of
the system (two spin-1/2 paricles), the energy transfer between the
system and the field happens, which is described by the relaxation
term $a^{\dagger}\sigma_i^-$ and the backaction term $a\sigma_i^+$.
This example is very different from Refs \cite{yu1, yu2}, in which
the two qubits couples with two independent environments separately.
In fact the couplings with the same field could induce an effective
interaction between two qubits, which manipulates the entanglement
between the two qubites \cite{braun}. Furthermore, as will be shown
in the following, the phenomenon of ESD is sensitive to the initial
conditions, and the energy transfer because of the interaction is
directly related to the disentanglement and revival of the
entanglement.

The evolution operator $U(t)=\exp(-iHt)$ can be calculated exactly
\cite{kim}. Choose the initial state $\rho(0)=(\frac{1-r}{4} +
r\ket{\varphi}\bra{\varphi})\bigotimes\ket{0}\bra{0}$, in which
$\ket{\varphi}=\sin\theta\ket{ee}+\cos\theta\ket{gg}$ with
$\ket{e(g)}$ denoting two eigenstates of spin-1/2 particles, and
$\ket{0}$ is the vacuum state for the quantized field. The
entanglement in the initial state is measured by the purity $r$ and
mixing $\theta$. Then the density matrix for the system
$\varrho(t)=\text{Tr}_E [U(t)\rho(0)U^{\dagger}(t)]$ can be
expressed in the basis $\{\ket{1}=\ket{ee}, \ket{2}=\ket{gg},
\ket{3}=\ket{eg}, \ket{4}=\ket{ge}\}$,
\begin{eqnarray}
\varrho_{11}&=&\frac{1-r}{4}+r(\sin\theta\frac{\cos\sqrt{6}gt+2}{3})^2,\nonumber\\
\varrho_{22}&=&\frac{1-r}{4}+r\cos^2\theta+2r(\sin\theta\frac{\cos\sqrt{6}gt-1}{3})^2.\nonumber\\
\varrho_{12}&=&\varrho_{21}=r\frac{\cos\sqrt{6}gt+2}{3}\sin\theta\cos\theta,\nonumber\\
\varrho_{33}&=&\varrho_{44}=\frac{1-r}{4}+ r
\frac{\sin^2\sqrt{6}gt\sin^2\theta}{6},\nonumber\\
\varrho_{34}&=&\varrho_{43}=r
\frac{\sin^2\sqrt{6}gt\sin^2\theta}{6}.
\end{eqnarray}

\begin{figure}
\begin{overpic}[width=6cm]{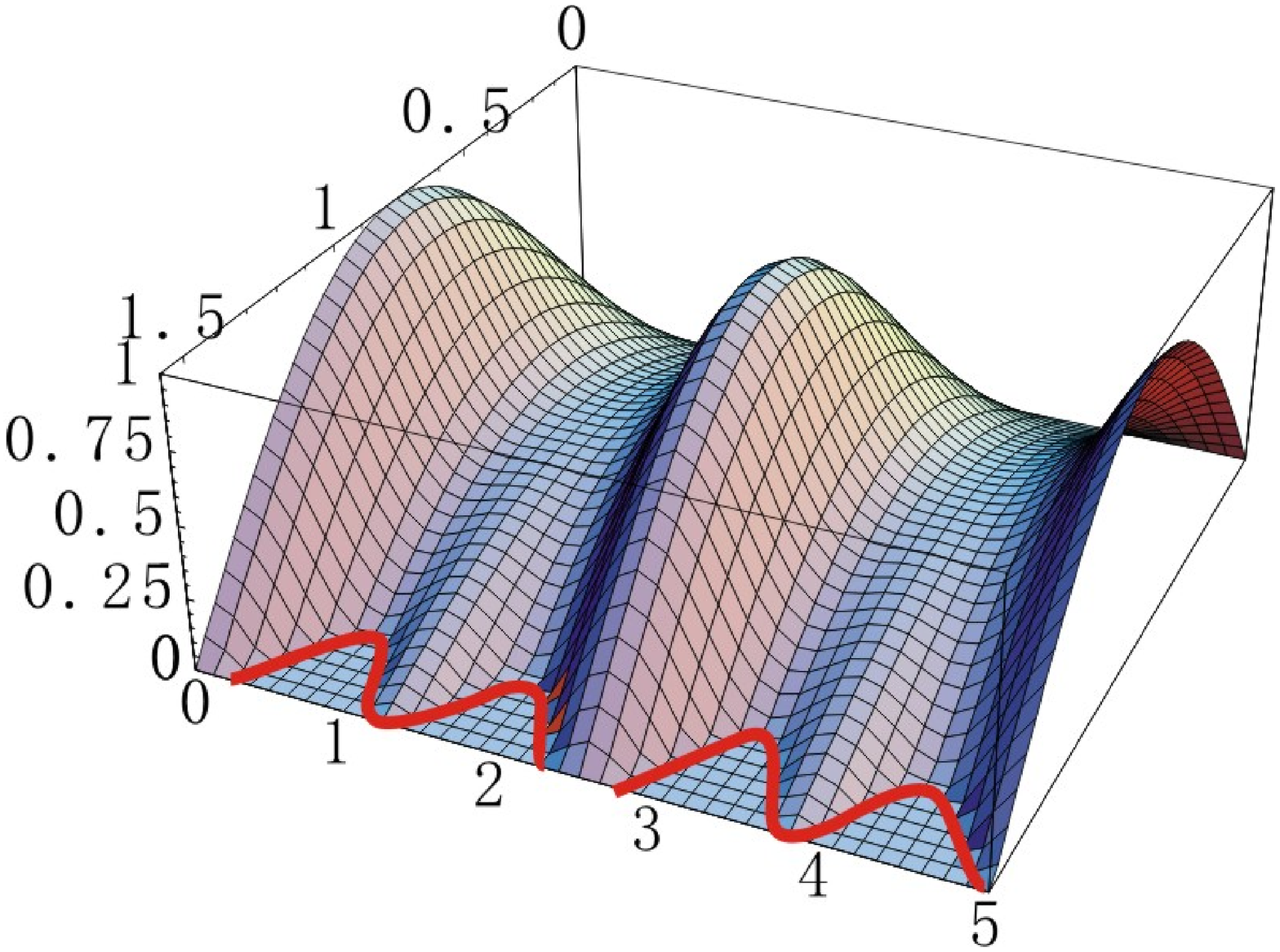}
\put(15,75){(a)} \put(40, 5){$gt$} \put(20, 65){$\theta$} \put(0,
25){\begin{rotate}{90}\text{concorrence}\end{rotate}}
\end{overpic}
\begin{overpic}[width=6cm]{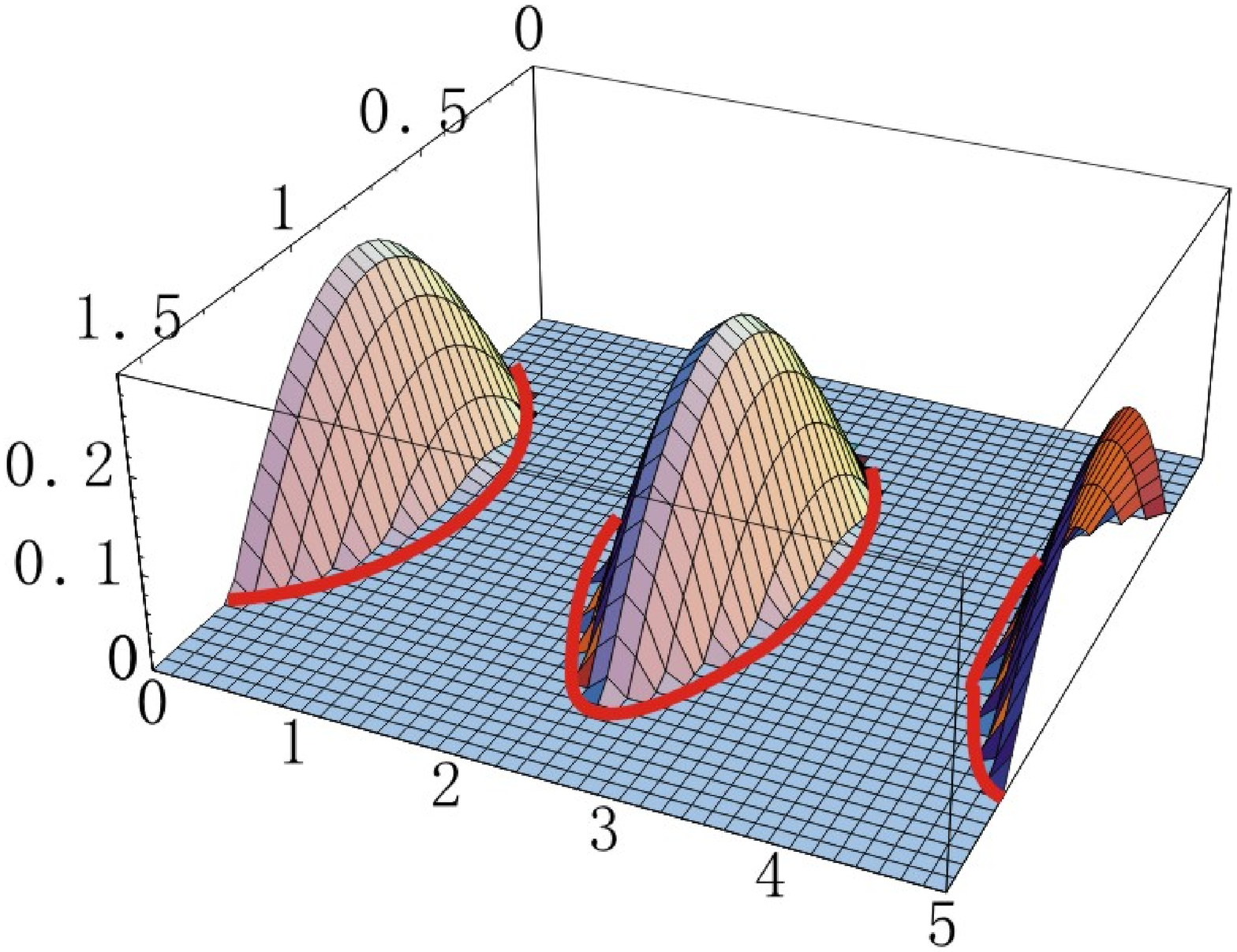}
\put(15,75){(b)} \put(40, 5){$gt$} \put(20, 65){$\theta$} \put(0,
25){\begin{rotate}{90}\text{concorrence}\end{rotate}}
\end{overpic}
\caption{\label{ca} The concurrence for $\varrho(t)$ versus the
coupling $gt$ and the mixing $\theta$. (a) corresponds to $r=1$ and
(b) for $r=0.5$. In both figures the points with vanishing
concurrence have been highlighted by red lines. }
\end{figure}

The concurrence $c=\max\{0,
\varrho_{34}-\sqrt{\varrho_{11}\rho_{22}}\}$ has been calculated, as
shown in Fig.\ref{ca}, for different initial states. Obviously the
concurrence is fluctuating with the rescaled time $gt$ and can be
zero in a finite time. We have highlight the points with red color
that the concurrence is vanishing, as the so-called entanglement
sudden death (ESD),  in Fig.\ref{ca} in this plot. An important
point is that ESD is sensitive to the initial state as displayed in
Fig.\ref{ca}. For the mixed initial state $r<1$,  the ESD happens
readily (Fig.\ref{ca}b). However, for the pure initial states, the
region with vanishing concurrence is compressed greatly
(Fig.\ref{ca}a). This phenomenon shows that ESD is completely
initial-state sensitivity in this model.

\begin{figure}[tpd]
\begin{overpic}[width=9cm]{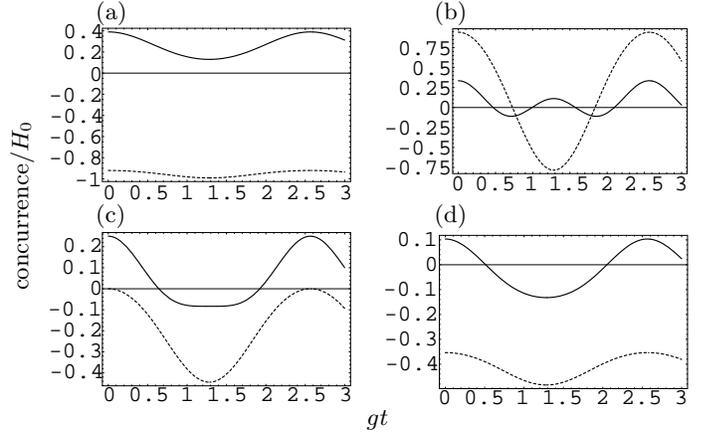}
\put(0, 20){\begin{rotate}{90}\text{concurrence}/$H_0$\end{rotate}}
\put(50, 0){$gt$} \put(10, 60){(a)} \put(60, 60){(b)} \put(10,
30){(c)} \put(60, 30){(d)}
\end{overpic}
\caption{\label{ce1} The concurrence (solid line) and $H_0$ (dashed
line) versus the rescaled time $gt$. For this plot we have chosen
(a)r=1, $\theta$=0.2; (b)r=1, $\theta$=1.4; (c)r=0.5,
$\theta$=$\pi$/4; (d)r=0.5, $\theta$=$\pi$/8. One should note that
we also plot the concurrence below $0$ and chose $\omega_0=1$ for
plotting $H_0$. }
\end{figure}

Another important character revealed by Fig.\ref{ca} is the
fluctuation/revival of entanglement. This phenomenon have been
discussed qualitatively in Ref. \cite{ficek}. However a clearer
interpretation is absent; How the entanglement is constructed by the
interaction and why and where the entanglement decreases with the
evolution? What is the physics behind the dynamics of entanglement?
In this part we try to give an enlightening discussion for this
problem based on this model. Let us first review the Hamiltonian Eq.
\eqref{hjc}; Obviously the dynamics of the two independent spin-1/2
particles is determined by the interaction terms in Eq. \eqref{hjc},
which is the charge of the energy transfer between the system and
field. Hence it is natural to check simultaneously the variation of
the energy $\langle H_0 \rangle_{\varrho}=
\langle\frac{\omega_0}{2}(\sigma_A^z +
\sigma_B^z)\rangle_{\varrho}=\varrho_{11}-\varrho_{22}$ and
concurrence in the system. The analytical relation between $c$ and
$\langle H_0 \rangle_{\varrho}$ is complicated and it is convenient
to plot for displaying their relation. One could find the variations
of $\langle H_0 \rangle_{\varrho}$ and concurrence is almost
in-step, as shown in Fig.\ref{ce1}. At most cases, with the
increment of $\langle H_0 \rangle_{\varrho}$, the concurrence
increases until the same value with the initial states. An
exceptional situation appears in Fig.\ref{ce1}(b), in which the
second wave peak accompanies with the opposite transfer of the
energy to the other situations. However we could suggest that the
dynamics of entanglement and its revival in this model stem from the
energy transfer between the system and environment. Moreover we also
note that the extreme points of the concurrence and those of
$\langle H_0 \rangle_{\varrho}$ are one-to-one. These phenomena show
the intimate relation between the concurrence and the energy
$\langle H_0 \rangle_{\varrho}$ in this model.

However the relation between ESD and the transfer of energy is not
direct, as shown in Fig.\ref{ce1}. Originally ESD comes from the
cutoff in the definition of concurrence. Although the physical
explanation for the dynamics of concurrence can be available, ESD
cannot be explained properly based on the transfer of energy in this
model. A further discussion is needed beyond this model.

\subsection{Dephasing Model}
In the previous subsection we discuss the dynamics of entanglement
and ESD in a dissipative two spin-1/2 system. A novel character is
the direct relation between the evolution of the concurrence and
energy transfer between the system and field. Another important
situation for the open systems is the dephasing case, in which there
is no energy transfer between the system and environment. A typical
Hamiltonian for this case can be written as
\begin{eqnarray}\label{hd}
H&=&H_S+H_E+H_I, \\
H_S&=&\frac{\omega_0}{2}(\sigma_A^z +
\sigma_B^z)+\Omega(\sigma_A^{\dagger}\sigma_B^-+\sigma_B^{\dagger}\sigma_A^-),\nonumber\\
H_E&=&\sum_j\omega_jb_j^{\dagger}b_j,\nonumber\\
H_I&=&(\sigma_A^{\dagger}\sigma_B^-+\sigma_B^{\dagger}\sigma_A^-)\sum_j\Gamma_j(b_j^{\dagger}+b_j),\nonumber
\end{eqnarray}
where two interacting qubits is coupling with a quantized
environment by $H_I$. The evolution operator can be given by,
\begin{eqnarray}\label{ub}
U(t)&=&\exp(-iHt)\nonumber\\
&=&e^{-it(\omega_0+\sum_jb_j^{\dagger}b_j)}\ket{ee}\bra{ee}+
e^{-it(-\omega_0+\sum_jb_j^{\dagger}b_j)}\ket{gg}\bra{gg}\nonumber\\
&&+e^{-it(\Omega-\sum_j\Gamma_j^2/\omega_j +
\sum_j\omega_ja_{j+}^{\dagger}a_{j+})}\ket{+}\bra{+}\nonumber\\
&&+e^{it(\Omega+\sum_j\Gamma_j^2/\omega_j -
\sum_j\omega_ja_{j-}^{\dagger}a_{j-})}\ket{-}\bra{-},
\end{eqnarray}
in which $\ket{g(e)}$ is the eigenstate of spin-1/2 particle and
$\ket{\pm}=(\ket{eg}\pm\ket{ge})/\sqrt{2}$ and
$a_{j\pm}=b_j\pm\Gamma_j/\omega_j$. Choose the initial state
$\rho(0)=(\frac{1-r}{4} +
r\ket{\phi}\bra{\phi})\bigotimes\ket{0}\bra{0}$, in which
$\ket{\phi}=\sin\theta\ket{eg}+\cos\theta\ket{ge}$, and at time $t$
the reduced density matrix
$\varrho(t)=\text{Tr}_E[U(t)\rho(0)U^{\dagger}(t)]$ can be written
in the basis $\{\ket{1}=\ket{ee}, \ket{2}=\ket{gg}, \ket{3}=\ket{+},
\ket{4}=\ket{-}\}$ as
\begin{eqnarray}\label{rho2}
\varrho_{11}&=&\varrho_{22}=\frac{1-r}{4},\nonumber\\
\varrho_{33}&=&\varrho_{44}+r\sin2\theta=\frac{1-r}{4}+\frac{r}{2}(1+\sin2\theta),\nonumber\\
\varrho_{34}&=&\varrho_{43}^*=- e^{-2i\Omega t}\prod_j
\exp(4\Gamma_j^2\frac{\cos\omega_jt-1}{\omega_j^2})\frac{r\cos2\theta}{2}.
\end{eqnarray}
Obviously the off-diagonal terms is damping because of the
decoherence factor $\prod_j
\exp[4\Omega_j^2(\cos\omega_jt-1)/\omega_j^2]$ \cite{sun}.

\begin{figure}[tbd]
\begin{overpic}{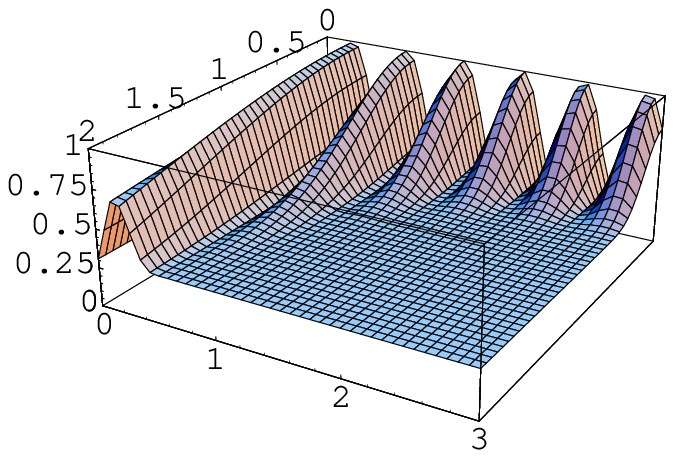}
\put(15,75){(a)} \put(40, 5){$\omega t$} \put(20,
65){$\Gamma/\omega$} \put(0,
25){\begin{rotate}{90}\text{concorrence}\end{rotate}}
\end{overpic}
\begin{overpic}[width=6cm]{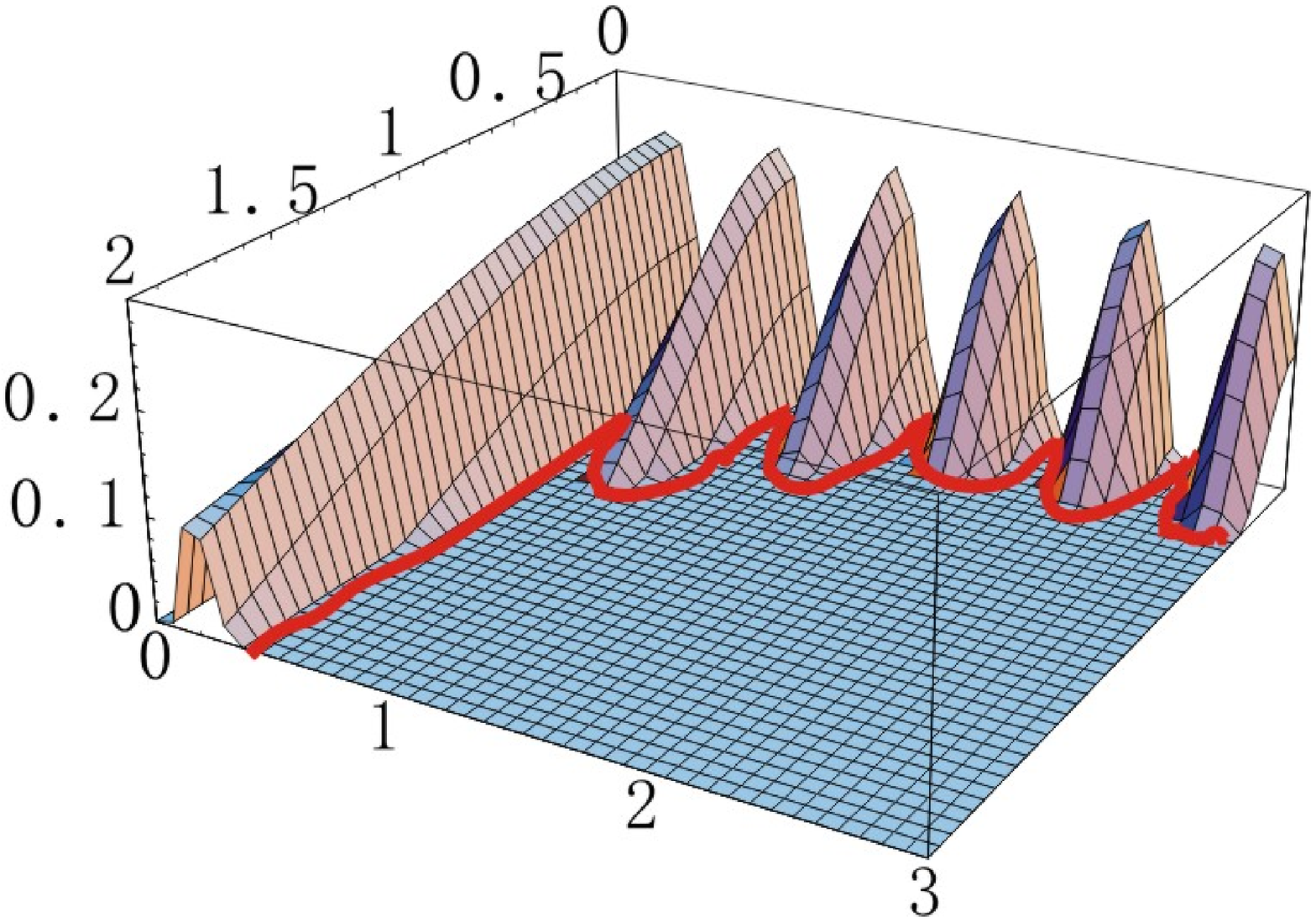}
\put(15,75){(b)} \put(40, 5){$\omega t$} \put(20,
65){$\Gamma/\omega$} \put(0,
25){\begin{rotate}{90}\text{concorrence}\end{rotate}}
\end{overpic}
\caption{\label{cb} The concurrence for $\varrho(t)$ versus the
rescaled time $\omega t$ and the coupling $\Gamma/\omega$. (a)
corresponds to $r=1$ and (b) for $r=0.5$. In both figures we choose
$\theta=\pi/20$ and $\Omega/\omega=3$. In (b) the red line
high-lightens the points of vanishing concurrence.}
\end{figure}

The concurrence for $\varrho(t)$ can be given exactly, as shown in
Fig.\ref{cb} with a single mode  for simplicity. One notes that the
entanglement does \emph{not} decrease, but have a damping
fluctuation until the same entanglement contained in the initial
state, as shown in Fig. \eqref{cb}(a). Another facet is that when
the concurrence is zero for the initial state, it is possible for
system to generate entanglement, the entanglement is also a
damping-oscillating function of time,  as shown in Fig.\ref{cb}(b).
The ESD can happen at some special times. However, for the entangled
initial states, there is no ESD at any time \cite{foot}.

\section{Entanglement in closed systems}
In the previous section, we discuss two models coupled with a field
or environment. In both models the dynamics of entanglement is
dominated by the coupling with the field or environment. Next we
will show that ESD can also appear in closed quantum systems, in
which the evolution of the system is insulated to its surroundings.
Let consider two spin-1/2 particles with Ising-type interaction. The
Hamiltonian is
\begin{equation}\label{ising}
H=\frac{\omega}{2}(\sigma_A^z+\sigma_B^z)+
\frac{g}{2}\sigma_A^x\sigma_B^x.
\end{equation}
The time evolution $U(t)=\exp(-iHt)$ is written in the basis
$\{\ket{1}=\ket{ee}, \ket{2}=\ket{gg}, \ket{3}=\ket{eg},
\ket{4}=\ket{ge}\}$,
\begin{eqnarray}
U_{11}&=&U^*_{22}=\cos\frac{\lambda t}{2}-
i\frac{2\omega}{\lambda }\sin\frac{\lambda t}{2},\nonumber\\
U_{12}&=&U_{21}=-i\frac{g}{\lambda}\sin\frac{\lambda_2t}{2},\nonumber\\
U_{33}&=&U_{44}=\cos\frac{gt}{2},\nonumber\\
U_{34}&=&U_{43}=-i\sin\frac{gt}{2},
\end{eqnarray}
where $\lambda=\sqrt{4\omega^2+g^2}$.

\begin{figure}[tbd]
\begin{overpic}[width=5cm]{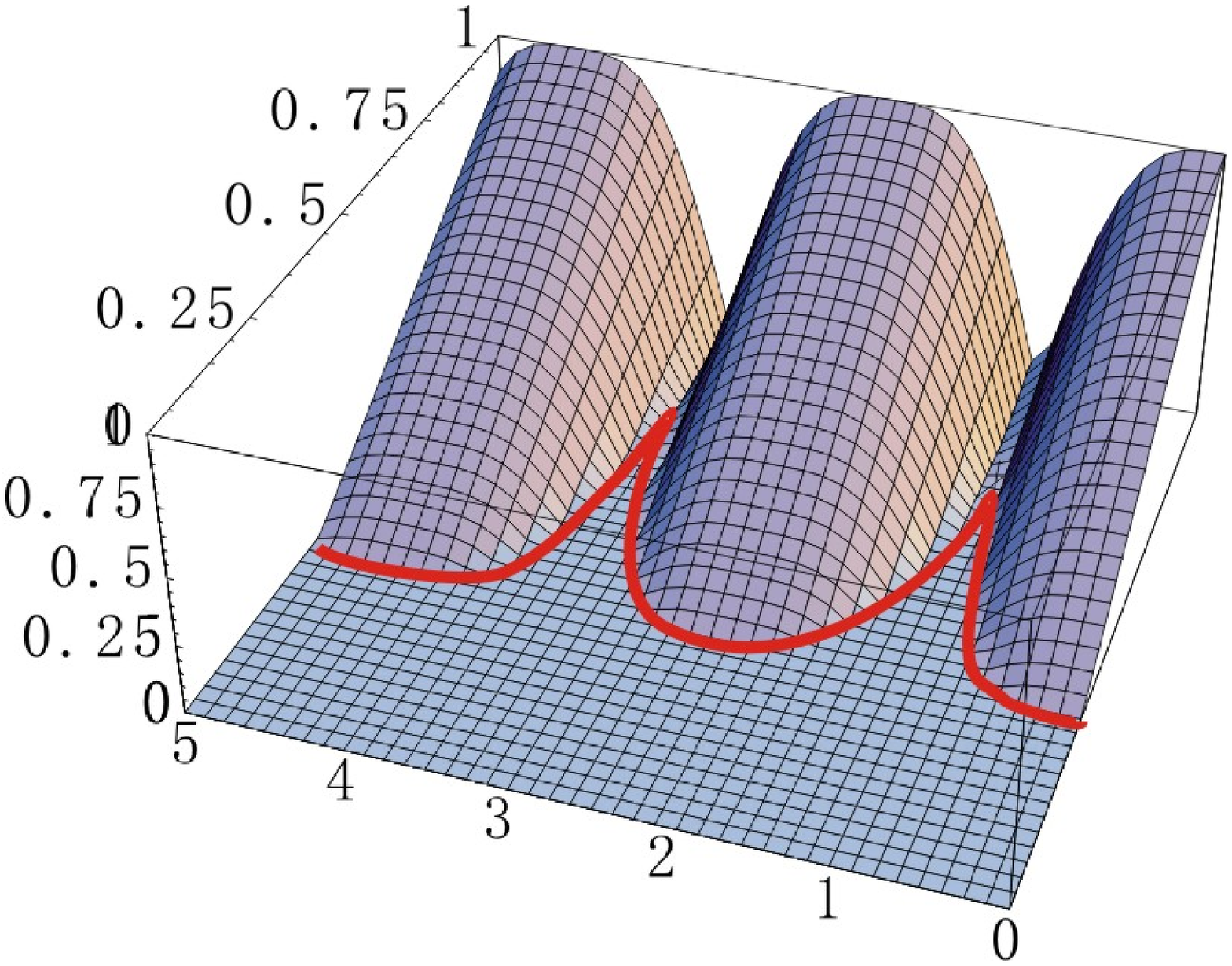}
\put(15,75){(a)} \put(40, 5){$\omega t$} \put(20, 65){$r$} \put(0,
25){\begin{rotate}{90}\text{concorrence}\end{rotate}}
\end{overpic}
\begin{overpic}[width=3cm]{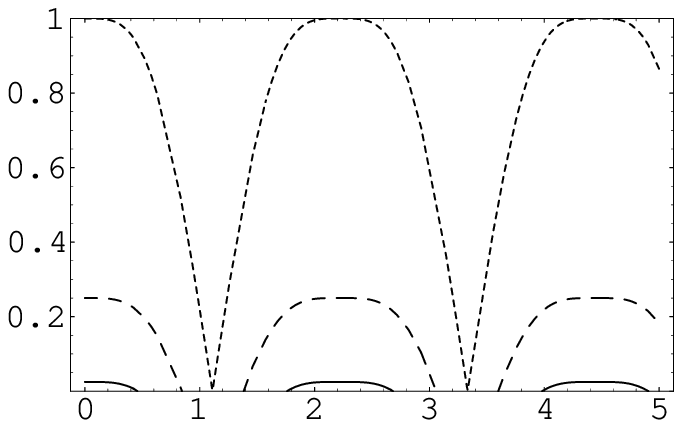}
\put(50, -5){$\omega t$} \put(0,
10){\begin{rotate}{90}\text{concorrence}\end{rotate}}
\end{overpic}
\begin{overpic}[width=5cm]{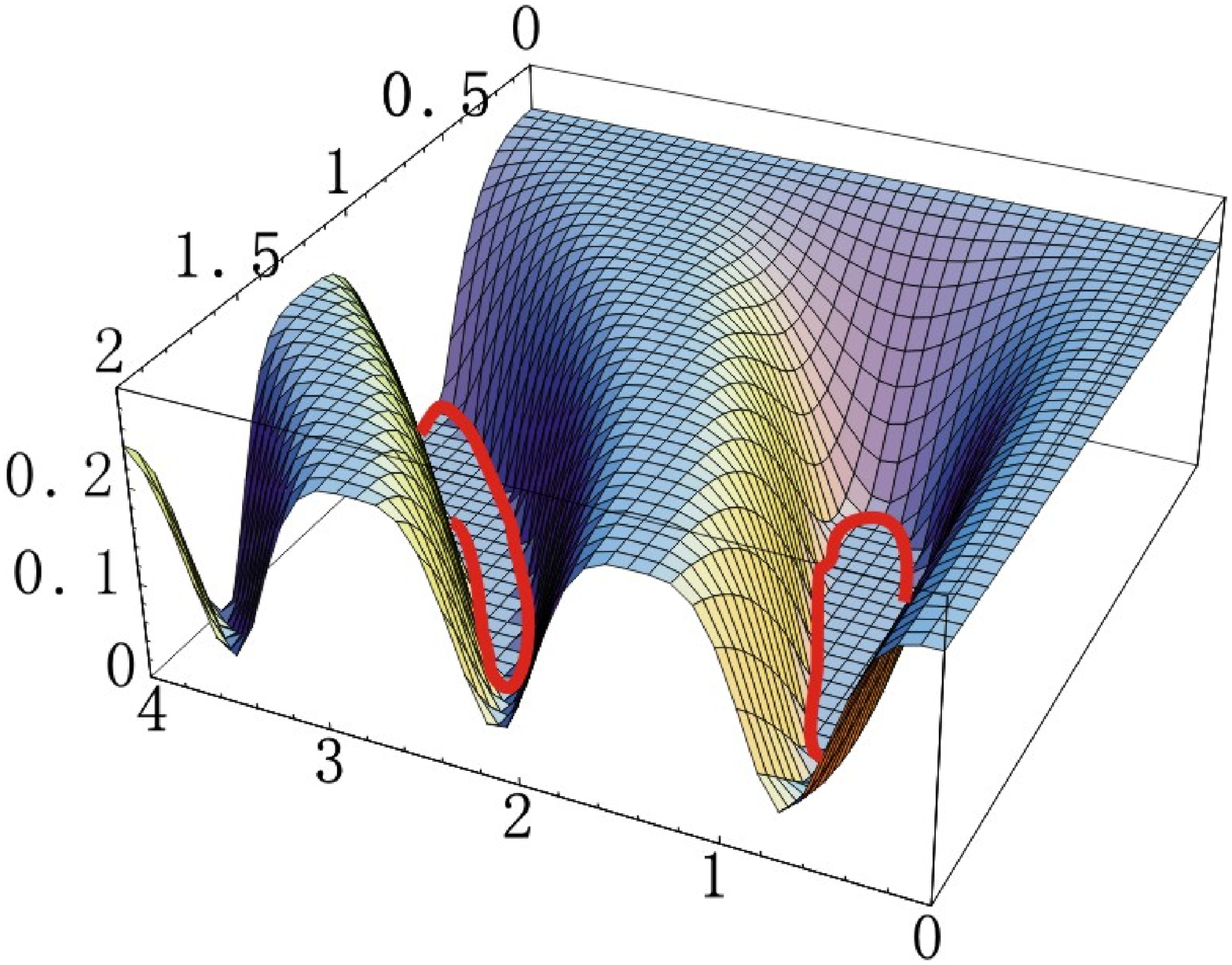}
\put(15,75){(b)} \put(40, 5){$\omega t$} \put(20, 65){$J$} \put(0,
25){\begin{rotate}{90}\text{concorrence}\end{rotate}}
\end{overpic}
\begin{overpic}[width=3cm]{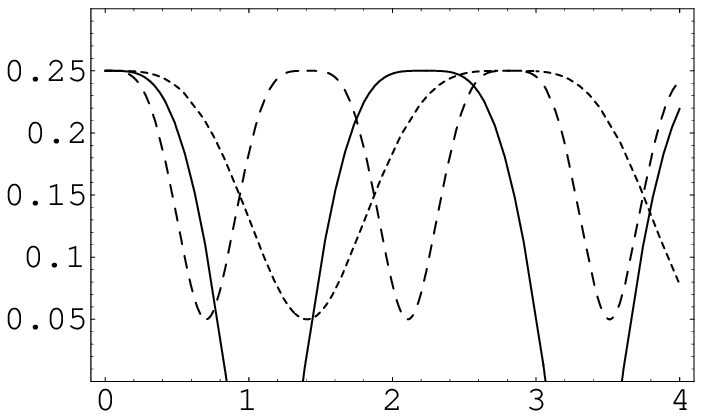}
\put(50, -5){$\omega t$} \put(0,
10){\begin{rotate}{90}\text{concorrence}\end{rotate}}
\end{overpic}
\caption{\label{cc} The concurrence versus the time $\omega t$ and
(a) the purity $r$, (b) the rescaled coupling $J=g/2\omega$. We have
chosen $\theta=\pi/4$ and (a) $J=1$, (b)$r=1/2$, and the points with
vanishing concurrence have been highlighting with red color. On the
right of the figure (a) and (b), the sectional drawings have also
provided with the same values of parameters respectively. The chosen
parameters for the solid, dotted and dashed lines are (a)$r=0.35, 1,
0.5$ and (b) $J=1, 0.5, 2$. }
\end{figure}

It is convenient to choose the initial state $\rho=\frac{1-r}{4} +
r\ket{\varphi}\bra{\varphi}$ with
$\ket{\varphi}=\sin\theta\ket{ee}+\cos\theta\ket{gg}$. The
entanglement for $\rho(t)=U(t)\rho U^{\dagger}(t)$ can be given
easily, and the concurrence have been plotted in Fig.\ref{cc}. It is
clear that ESD happens for some special cases and then the
entanglement revives at a later time. We also choose the initial
state $\rho=\frac{1-r}{4} + r\ket{\phi}\bra{\phi}$ with
$\ket{\phi}=\sin\theta\ket{eg}+\cos\theta\ket{ge}$. However there is
no ESD . These phenomenon show again that ESD is sensitive to the
initial conditions.

\begin{figure}[tpd]
\begin{overpic}[width=9cm]{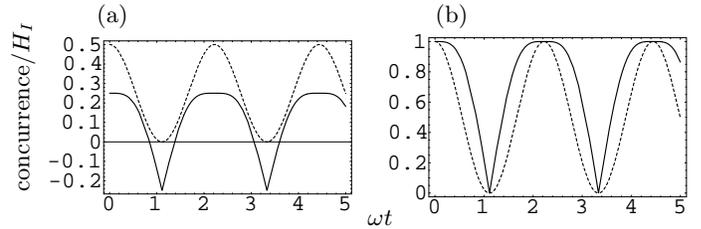}
\put(0, 5){\begin{rotate}{90}\text{concurrence}/$H_I$\end{rotate}}
\put(50, 0){$\omega t$} \put(10, 30){(a)} \put(60, 30){(b)}
\end{overpic}
\caption{\label{ce3} The concurrence (solid line) and $H_I$ (dahsed
line) versus the time $\omega t$. For this plot we have chosen
$\theta=\pi/4$ and $J=1$. The figure (a) corresponds $r=0.5$ and (b)
for $r=1$. }
\end{figure}

It is of great interest to check the relation between the
concurrence and the energy transfer described by $\langle H_I
\rangle_{\rho}=\langle g\sigma_A^x\sigma_B^x\rangle_{\rho}$. It is
easily to note that the concurrenc $c$ of $\rho(t)$ is $c\propto
\langle H_I \rangle_{\rho}/J$ and a figure has been drawn for
showing their relation , displayed in Fig.\ref{ce3}. It is very
interesting to note that the evolution of concurrence and $\langle
H_I \rangle_{\rho}$ is in-step and there is a one-to-one relation of
the extreme points between the concurrence and $\langle H_I
\rangle_{\rho}$. This phenomenon is similar to the case discussed in
Sec.IIA (see Fig.\ref{ce1}) and show that the dynamics of
entanglement can be related directly to the energy transfer in this
model.

\section{discussions and conclusions}
Some conclusions and further discussions should be presented in this
section. In this paper we discussed the dynamics of entanglement and
ESD in bipartite systems . We found two very different types of the
dynamics of entanglement. The first case is that there is the effect
of the dissipation and vice versa, just as shown in Eq. \eqref{hjc}
and \eqref{ising}. Our calculation shows the intimate relation
between the concurrence and the energy transfer described by  $H_0$
in Eq. \eqref{hjc} and $H_I$ in Eq. \eqref{ising}; the extreme
points of concurrence and the energy of $H_0$ or $H_I$ under the
time evolution are one-to-one. In the case without ESD, the minimums
of the energy corresponds to that of the concurrence , as shown in
Fig. \eqref{ce1}(a) and Fig. \eqref{ce3}(b). However, when ESD
happens, the correspondence is destroyed.

In fact mathematically ESD stems from the cutoff in the definition
of concurrence, and so it is difficult to find a physical
interpretation. We suggest that the concurrence could be expressed
as $c=\max\{0, f(E)\}$, in which $f(E)$ is a function of the energy
transfer, e.g. $f(E)\propto\langle H_I \rangle_{\rho}/J$ in Sec.
III. Then we could define a critical energy $E_c$, below which the
system must be non-entangled , and above which the system must be
entangled. When $E_c$ is the minimal of the energy of the system,
then we immediately conclude there is no ESD. Contrarily ESD could
happen. Unfortunately the critical energy may be complicated and be
dependent on the state as shown in Fig.\ref{ce1} and \ref{ce3} and
prevent us from giving a further analysis.

Another interesting case is the dephasing model, which has been
discussed extensively in the past as one of the standard decoherence
models \cite{preskill}. The evolution of concurrence in the model
Eq. \eqref{hd} is damping-oscillating with time until the same
entanglement contained in the initial state (see Fig. \ref{cb}).
Moreover the oscillation of concurrence seems  not directly related
to the energy transfer since there is no energy transfer between
system and environment. From this phenomenon it may be frustrating
that the connection of concurrence and the energy is not universal.
What are the reasons of the dynamics of concurrence in the dephasing
case is still an open question. The phenomenon of ESD does not
appear in the dephasing model when the initial state is entangled.
It seems to imply that ESD intrinsically originates from the energy
transfer. However the correct understanding depends on the physical
meanings of the concurrence.

In conclusion we have discussed the dynamics of the entanglement and
ESD in some special models. Some interesting  phenomena have been
presented with enlightening discussions. In our own points the main
obstacle of explaining these phenomena is that it is unclear what
the physical meanings of the concurrence are. Recently a great deal
of works are denoted to the understanding from the energy of the
system \cite{toth}. However, as shown in our paper, it maybe fail
when there is ESD happening. More recently a  physical
interpretation of concurrence for the bipartite systems  has been
provided based on the Casimir operator in \cite{Klyachko}. It maybe
opens another door to understand concurrence as a physical quantity.

This work was supported by NSF of China under grants 10305002 and
60578014.


\begin{thebibliography}{99}
\bibitem{foot}The paper \cite{roszak} discussed the disentanglement in a
dephasing model with thermal bath , and the authors show that ESD
happens only in a high temperature with special initial state. The
result is not contratry to our discussion since our system is
embedded in zero-temperature enviroment.
\end{thebibliography}
\end{document}